\documentclass[12pt]{article}
\pdfoutput=1
\usepackage{epsf,amsfonts,amssymb,epsfig,amsmath,graphics,slashed,cite}
\usepackage{hyperref,graphicx, subfig}
\hypersetup{colorlinks=true}
\newcommand{\nn}{\notag \\}

\begin{document}

\makeatletter
\renewcommand{\theequation}{\thesection.\arabic{equation}}
\@addtoreset{equation}{section}
\makeatother

\baselineskip 18pt

\begin{titlepage}

\vfill

\begin{flushright}
Imperial/TP/2013/JG/01\\
\end{flushright}

\vfill

\begin{center}
   \baselineskip=16pt
   {\Large\bf  Holographic charge density waves}
  \vskip 1.5cm
      Aristomenis Donos and Jerome P. Gauntlett\\
   \vskip .6cm
      \begin{small}
      \textit{Blackett Laboratory, Imperial College\\
        Prince Consort Rd., London, SW7 2AZ, U.K.}
        \end{small}\\*[.6cm]

\end{center}

\vfill

\begin{center}
\textbf{Abstract}
\end{center}

\begin{quote}
We show that strongly coupled holographic matter at finite charge density can exhibit charge density wave phases
which spontaneously break translation invariance while preserving time-reversal and parity  invariance.
We show that such phases are possible within Einstein-Maxwell-dilaton theory in general spacetime dimensions.
We also discuss related spatially modulated phases when there is an additional coupling to a 
second vector field, possibly with non-zero mass. We discuss how these constructions, and others, 
should be associated with novel spatially modulated ground states.
\end{quote}

\vfill

\end{titlepage}
\setcounter{equation}{0}


\section{Introduction}

The charge density in a metal is usually highly uniform due to strong Coulombic interactions. However, a wide variety of metals exhibit
stable charge density wave (CDW) phases in which the charge density spontaneously becomes spatially modulated.
These phases were first predicted by Peierls in the context of weakly coupled one-dimensional systems \cite{peierls}.
They are also known to occur in strongly correlated systems and in some cases the CDWs are present with additional order. 
For example, in the high temperature cuprate superconductors
there are striped phases in which CDWs appear with spin density waves, with the latter breaking time-reversal invariance (for a review see \cite{vojta}).
In this paper we will discuss CDWs, that preserve time-reversal and parity invariance ($T$ and $P$),
within the holographic framework of the AdS/CFT correspondence.

The first constructions of holographic spatially modulated phases, but without CDWs, 
were made in the context of $D=5$ Einstein-Maxwell theory
with a Chern-Simons term \cite{Nakamura:2009tf,Donos:2012wi} (see also \cite{Domokos:2007kt}). 
At finite charge density and high temperatures, the system is described by the standard electrically charged AdS-RN black brane
corresponding to a spatially homogeneous and isotropic phase. For sufficiently large Chern-Simons coupling, the AdS-RN black brane becomes unstable at a critical temperature and a new branch of black hole solutions appears corresponding, in the dual field theory,
to a phase with spatially modulated currents with helical order \cite{Nakamura:2009tf}. 
The fully back-reacted 
black hole solutions were constructed in \cite{Donos:2012wi} where it was shown that helical current phase is thermodynamically
preferred and that the phase transition is second order. Furthermore,
at zero temperature 
the black hole solutions of  \cite{Donos:2012wi} revealed new ground states with a helical structure similar to those studied in
\cite{Iizuka:2012iv}. 
These helical current phases break $P$ and $T$.

In subsequent work, a $D=4$ Einstein-Maxwell theory coupled to a pseudo-scalar $\varphi$  was shown to admit black hole solutions corresponding to phases with spatially modulated currents and in addition, CDWs  \cite{Donos:2011bh}.
In this class of models the key coupling driving the spatially modulated phase transition is the axion-like coupling $\varphi F\wedge F$, where $F$ is
the field strength of the Maxwell field. The fully
back-reacted black hole solutions require solving partial differential equations and some results have 
recently appeared in \cite{Rozali:2012es}. While the models of \cite{Donos:2011bh} realise CDWs the spatially modulated
currents break $P$ and $T$. Holographic constructions of spatially modulated phases at finite charge density\footnote{Spatially modulated phases preserving the 
$U(1)$ symmetry have also been shown to exist in the presence of magnetic fields in 
\cite{Donos:2011qt,Donos:2011pn,Jokela:2012vn,Alsup:2012kr,Cremonini:2012ir,Jokela:2012se,Donos:2012yu},
where the magnetic field itself breaks $T$.}  
have also been made in \cite{Ooguri:2010kt,Ooguri:2010xs,Bayona:2011ab,Bergman:2011rf,Takeuchi:2011uk,Donos:2011ff,Iizuka:2012pn,Iizuka:2013ag} 
and $P$ and $T$ are again not preserved.

It is natural to ask, therefore, if breaking of $P$ and/or $T$ is necessary to realise a spatially modulated phase at finite
charge density in the context of holography. Here we will show that it is not\footnote{For a discussion of constructions exploiting magnetic gaugings, see 
\cite{Sachdev:2012tj}.}. Specifically, we will discuss two general classes of models associated with 
CDW phases, without current density waves, and preserving $P$ and $T$.
Our general strategy will be to consider set-ups in which
the zero temperature limit of the electrically charged black holes that describe the unbroken high temperature phase, 
are domain wall solutions interpolating between some UV fixed point and an 
electrically charged $AdS_2 \times\mathbb{R}^{D-2}$ solution in the far IR. We then construct CDW type modes that violate the $AdS_2$ BF bound; the existence of these modes
necessarily implies that the finite temperature unbroken-phase black holes become unstable at some critical temperature. 
Furthermore,
at this critical temperature the unbroken phase black holes will admit a static normalisable mode
corresponding to the existence of a new branch of electrically charged black holes dual to the CDW phase.

The first class of models, discussed in section 2, involve Einstein-Maxwell theory coupled to a single 
scalar field $\phi$, often called the dilaton. These models have a single conserved $U(1)$ symmetry in the dual field theory.
We will see that the relevant CDW mode involves the gauge-field, the dilaton and the metric.

The second class of models, discussed in section 3, involves Einstein-Maxwell-dilaton models
with an additional coupling to a second vector field which can either be massive or massless.
 This class of models is somewhat simpler than the first in that the standard electrically charged AdS-RN black brane
 solution solves the equations of motion. The AdS-RN black branes 
 describe the high temperature unbroken phase of a dual CFT when held at finite charge density with respect to  the $U(1)$ symmetry associated with the Maxwell field. When the second vector field is massless the
dual theory has a second global $U(1)$ symmetry. We show that there can be spatially modulated modes just
involving the dilaton and the second vector field, which correspond, in the massless case, to CDWs for the second $U(1)$ symmetry.

In section 4 we conclude with some discussion on how our constructions should be associated with novel spatially
modulated ground states. One route is to follow the spatially modulated phases down to zero temperature. 
A different route, following \cite{Donos:2012js}, 
utilises $AdS_2$ solutions which contain spatially modulated modes that are dual to relevant operators in the RG sense.

\section{Einstein-Maxwell-Dilaton models}
In this section we consider Einstein-Maxwell-dilaton models in $D$ spacetime dimensions, 
which couple gravity to a gauge-field $A$, with field strength $F=dA$, 
and a scalar field $\phi$, the ``dilaton". 
The Lagrangian density is given by
\begin{align}\label{eq:action}
\mathcal{L}=R-V\left(\phi\right)-\tfrac{1}{2}(\partial\phi)^2-\tfrac{1}{4}\,\tau\left(\phi\right)F^2\,,
\end{align}
where $F^2=F_{\mu\nu}F^{\mu\nu}$ and $V,\tau$ are functions that we will partially restrict below.
The equations of motion derived from the action \eqref{eq:action} can be written as
\begin{align}\label{eq:eom}
&R_{\mu\nu}=\tfrac{1}{D-2}V\,g_{\mu\nu}+\tfrac{1}{2}\partial_{\mu}\phi\,\partial_{\nu}\phi
+\tfrac{1}{2}\tau \left(F_{\mu\rho}F_{\nu}{}^{\rho}-\tfrac{1}{2\left(D-2\right)}g_{\mu\nu}F_{\rho\sigma}F^{\rho\sigma} \right)\,,\nn
&\nabla_\mu\left(\tau\, F^{\mu\nu} \right)=0\, ,\nn
&\nabla^2\phi-V^{\prime}-\tfrac{1}{4}\tau^{\prime}F^2=0\, .
\end{align}
This class of models has been studied in a holographic context in \cite{Charmousis:2010zz} (see also e.g. 
\cite{Taylor:2008tg,Goldstein:2009cv,Gubser:2009qt,Cadoni:2009xm,
Lee:2010qs,Perlmutter:2010qu,Cadoni:2011kv,Iizuka:2011hg,Ogawa:2011bz,Huijse:2011ef,Anantua:2012nj}).

We will consider particular classes of models that admit electrically charged $AdS_2\times\mathbb{R}^{D-2}$ solutions 
of the form
\begin{align}\label{eq:ads2}
ds^{2}&=L^{2}\,\left(ds^{2}\left(AdS_{2}\right)+dx_{1}^{2}+\dots+dx^{2}_{D-2}\right)\,,\nn
F&=E\,\mathrm{Vol}\left(AdS_{2}\right)\,,\nn
\phi&=\phi_0\,,
\end{align}
where $E, \phi_0, L$ are constants and
we have written the metric and field strength using two dimensional anti de-Sitter space with unit radius. 
Using the equations of motion \eqref{eq:eom} the existence of this class of solutions
imposes the conditions
\begin{align}\label{eq:req}
&V\left(\varphi_{0}\right)\,\tau^{\prime}\left(\varphi_{0}\right)=-\tau\left(\varphi_{0}\right)\,V^{\prime}\left(\varphi_{0}\right)\,,\nn
&V\left(\varphi_{0}\right)<0,\quad \tau\left(\varphi_{0}\right)>0\,,
\end{align}
and
\begin{align}\label{lsqe}
L^{2}=-\frac{1}{V\left(\varphi_{0}\right)},\quad E^2=\frac{2}{-V\left(\varphi_{0}\right)\tau\left(\varphi_{0}\right)}\,.
\end{align}
We will take $E>0$.
We will need the first few terms in the expansion of the potential $V$ and the function $\tau$ around the value $\varphi_{0}$. 
In general, it can be brought to the convenient form
\begin{align}\label{eq:pott_exp}
V&=v_{0}\left(1-\tau_{1} \,\left(\varphi-\varphi_{0}\right)-\frac{v_2}{2}\,\left(\varphi-\varphi_{0}\right)^{2}+\cdots\right)\,,\nn
\tau&=\tau_{0}\left(1+\tau_{1}\,\left(\varphi-\varphi_{0}\right)-\frac{\tau_{2}}{2}\,\left(\varphi-\varphi_{0}\right)^{2}+\cdots\right)\,,\nn
&v_{0}<0,\quad \tau_{0}>0\,,
\end{align}
where we have incorporated the conditions in \eqref{eq:req}.

We now wish to study linearized perturbations around the background \eqref{eq:ads2}. For simplicity we will
continue with $D=4$ but we will also quote the final key result for $D=5$.
More specifically, using a coordinate system for $AdS_{2}$ such that
\begin{align}\label{ads2coord}
ds^{2}\left(AdS_{2}\right)=-r^{2}\,dt^{2}+\frac{dr^{2}}{r^{2}}\,,
\end{align}
we are interested in the perturbation 
\begin{align}\label{eq:pert}
\delta g_{tt}&=L^{2}r^{2}\,e^{-i\omega t} h_{tt}\left(r\right)\,\cos\left(k\,x_{1}\right)\,,\nn
\delta g_{x_{i}x_{i}}&=L^{2}\,e^{-i\omega t} h_{x_{i}x_{i}}\left(r\right)\,\cos\left(k\,x_{1}\right)\,,\nn
\delta g_{tx_{1}}&=L^{2}\,e^{-i\omega t} h_{tx_{1}}\left(r\right)\,\sin\left(k\,x_{1}\right)\,,\nn
\delta A_{t}&=-E\,e^{-i\omega t} a_{t}\left(r\right)\,\cos\left(k\,x_{1}\right)\,,\nn
\delta A_{x_{1}}&=-E\,e^{-i\omega t} a_{x_{1}}\left(r\right)\,\sin\left(k\,x_{1}\right)\,,\nn
\delta \varphi&=e^{-i\omega t} h\left(r\right)\,\cos\left(k\,x_{1}\right)\,,
\end{align}
with $i=1,2$, which involves seven functions of the radius $\left\{h_{tt},h_{x_{i}x_{i}}, h_{tx_{1}},a_{t},a_{x_{1}},h\right\}$ in
a self consistent manner, as we shall see. 

The linearised perturbation is in a radial gauge but
there is still some remaining gauge freedom. Specifically, 
the combined co-ordinate transformation $x^{\mu}\rightarrow x^{\mu}+\delta x^{\mu}$
and $U(1)$ gauge transformation $A_\mu\rightarrow A_\mu+\partial_\mu\delta\Lambda$ with
\begin{align}\label{eq:rem_gauge}
\delta t&=-\frac{1}{2r^{2}}\,\partial_{t}f+g_{1}\,,\nn
\delta r&=r\,f\,,\nn
\delta x_{1}&=-\ln r\,\partial_{x_{1}}f+g_{2}\,,\nn
\delta x_{2}&=0\,,\nn
\delta\Lambda&=-\frac{E}{r}\,\partial_{t}f+E\,g_{3}\,,
\end{align}
keeps the form of the perturbation \eqref{eq:pert} invariant when $f=f(t,x_{1})$ and $g_{i}=g_{i}(t,x_{1})$. 
Indeed for such gauge transformations we have
\begin{align}\label{eq:gauge_tr}
\delta g_{tt}&=-2L^{2}\,\left( r^{2}f-\frac{1}{2}\,\partial_{t}^{2}f+r^{2}\partial_{t}g_{1}\right)\,,\nn
\delta g_{x_{1}x_{1}}&=2L^{2}\,\left(-\ln r\,\partial_{x_{1}}^{2}f+\partial_{x_{1}}g_{2} \right)\,,\nn
\delta g_{x_{2}x_{2}}&=0\,,\nn
\delta g_{tx_{1}}&=L^{2}\,\left(\frac{1}{2}\partial_{t}\partial_{x_{1}}f -r^{2}\,\partial_{x_{1}}g_{1}-\ln r\,\partial_{t}\partial_{x_{1}}f+\partial_{t}g_{2}\right)\,,\nn
\delta A_{t}&=E\,\left(-rf-\frac{1}{2}\frac{1}{r}\partial_{t}^{2}f-r\partial_{t}g_{1}+\partial_{t}g_{3}\right)\,,\nn
\delta A_{x_{1}}&=E\,\left(-\frac{1}{2r}\partial_{t}\partial_{x_{1}}f+\partial_{x_{1}}g_{3}-r\,\partial_{x_{1}}g_{1} \right)\,,
\end{align}
and we see that two gauge invariant combinations are given by
\begin{align}
\frac{1}{L^2 r^2}\delta g_{tt} -\frac{2}{E}\,\partial_{r}\delta A_{t}, \qquad \delta g_{x_{2}x_{2}}\,.
\end{align}
This will be useful shortly.

We now return to the linearised perturbation \eqref{eq:pert} about the $AdS_2\times\mathbb{R}^2$ solution
 \eqref{eq:ads2}. After substituting into the equations of motion
\eqref{eq:eom} we obtain three ordinary differential equations from the gauge field equation of motion, 
 seven from Einstein's equations and one from the scalar equation of motion:
\begin{align}\label{eq:orig_odes}
&-2 r^2 a_t'+r^3 (g_{x_{1}x_{1}}'+g_{x_{2}x_{2}}')+\omega ^2( g_{x_{1}x_{1}}+g_{x_{2}x_{2}})-2 i k \omega g_{tx_{1}}-r^4 g_{tt}'' \nn
&\qquad\qquad\qquad\qquad\qquad\qquad\qquad\qquad\qquad\qquad-3 r^3 g_{tt}'+k^2 r^2 g_{tt}-r^2 g_{tt}=0\,,\nn
&-i \omega  \left(-r \left(g_{x_{1}x_{1}}'+g_{x_{2}x_{2}}'\right)+g_{x_{1}x_{1}}+g_{x_{2}x_{2}}\right)+k r g_{tx_{1}}'-2 k g_{tx_{1}}=0\,,\nn
&2 a_t'-r \left(r \left(g_{x_{1}x_{1}}''+g_{x_{2}x_{2}}''-g_{tt}''\right)+g_{x_{1}x_{1}}'+g_{x_{2}x_{2}}'-3 g_{tt}'\right)+g_{tt}=0\,,\nn
&r^2 \left( 2a_t'+2r g_{x_{1}x_{1}}'+2\tau_{1} h+r^2 g_{x_{1}x_{1}}''+(1+k^2) g_{tt}-k^2g_{x_2x_2}\right)+\omega ^2 g_{x_{1}x_{1}}\nn
&\qquad\qquad\qquad\qquad\qquad\qquad\qquad\qquad\qquad\qquad\qquad\qquad\quad-2 i k \omega  g_{tx_{1}}=0\,,\nn
%
&\omega^2g_{x_{2}x_{2}} +r^2 \left(2a_t'+2r g_{x_{2}x_{2}}'+2\tau_{1} h-k^2g_{x_{2}x_{2}}+r^2 g_{x_{2}x_{2}}''+g_{tt}\right)=0\,,\nn
%
&r^2 \left(2\tau_{1} a_t'+r^2 h''+2 r h'-k^2 h-(\tau_{2}+v_{2})h+\tau_{1} g_{tt}\right)+\omega^2 h=0\,,\nn
%
&i \omega  \left(2 a_t'+g_{x_{1}x_{1}}+g_{x_{2}x_{2}}+g_{tt}+2 \tau_{1} h\right)+2kr^2a_{x_{1}}'+2k g_{tx_{1}}=0\,,\nn
&2 i k \omega  a_{x_{1}}-2 k^2 a_t+r^2 \left(2 a_t''+g_{x_{1}x_{1}}'+g_{x_{2}x_{2}}'+g_{tt}'+2 \tau_{1} h'\right)=0\,,\nn
&r^2 \left(2 a_{x_{1}}'+g_{tx_{1}}''\right)+i k \omega  g_{x_{2}x_{2}}=0\,,\nn
&-k \left(2 a_t+r \left(r \left(g_{tt}'-g_{x_{2}x_{2}}'\right)+g_{tt}\right)\right)+i \omega  \left(2 a_{x_{1}}+g_{tx_{1}}'\right)=0\,,\nn
&r^2 \left(r^2 a_{x_{1}}''+2 r a_{x_{1}}'+g_{tx_{1}}'\right)+\omega ^2 a_{x_{1}}+i k \omega  a_t=0\,.
\end{align}
Some analysis now shows that four of these ODEs are implied by the remaining seven, as expected from the
Bianchi identity for the Einstein tensor. Furthermore, we find that in this system
of seven ODEs the
three functions $\left\{h,a_{t},h_{x_{2}x_{2}} \right\}$ enter with second order derivatives in $r$ while the four functions 
$\left\{h_{x_{1}x_{1}},h_{tx_{1}},h_{tt},a_{x_{1}} \right\}$ enter with first order derivatives. 
This indicates that there are three propagating degrees of freedom and four constraints. 

To proceed we now exploit our previous discussion on gauge invariance and introduce the variables
\begin{align}\label{eq:AdS2_sc_def}
\Phi_{1}=h_{tt}+2 \,a_{t}^{\prime},\qquad \Phi_{2}=g_{x_{2}x_{2}},\qquad \Phi_3=h\,,
\end{align}
to find that the system of seven ODEs can be written in terms of $\{\Phi_1,\Phi_2,\Phi_3\}$ and
$\{h_{x_{1}x_{1}},h_{tx_{1}},h_{tt},a_{x_{1}}\}$
in the form:
\begin{align}\label{eq:reqns}
&\omega^2\Phi_3+r^3(r\Phi_3''+2\Phi_3')+r^2\left(\tau_1\Phi_1-(k^2+v_2+\tau_2)\Phi_3\right)=0\,,\nn
%
&\omega^2\Phi_2+r^3(r\Phi_2''+2\Phi_2')+r^2\left( \Phi_1-k^2\Phi_2+2\tau_1\Phi_3  \right)
=0\,,\nn
&2 i k  \omega ( r a_{x_{1}}+g_{tx_{1}})+2 k^2 r^2 a_t'-2 k^2 r a_t-\omega ^2 g_{x_{1}x_{1}}
+2 r^3 \tau_{1} \Phi_3' -r^2 (2 \tau_{1} \Phi_3+\Phi_{1})  \nn
&\qquad\qquad\qquad\qquad\qquad\qquad\qquad\qquad\qquad
-k^2 r^2 \Phi_{1}+k^2 r^2 \Phi_{2}+r^3 \Phi_{1}'-\omega ^2 \Phi_{2}=0\,,\nn
&-2 k^2 r^2 a_t'+r^3 g_{x_{1}x_{1}}'+\omega^2 g_{x_{1}x_{1}}-2 i k \omega  g_{tx_{1}}+2 r^2 \tau_{1} \Phi_3+\left(k^2+1\right) r^2 \Phi_{1}\nn
&\qquad\qquad\qquad\qquad\qquad\qquad\qquad\qquad\qquad+ \left(\omega ^2-k^2 r^2\right)\Phi_{2}+r^3 \Phi_{2}'=0\,,\nn
%
&i \omega  \left(-2 k^2 r^2 a_t'+2 r^2 \tau_{1} \Phi_3+\left(k^2+1\right) r^2 \Phi_{1}-k^2 r^2 \Phi_{2}+r^2 \Phi_{2}+\omega ^2 \Phi_{2}+\left(r^2+\omega ^2\right)g_{x_{1}x_{1}}\right)\nn
&\qquad\qquad \qquad\qquad  \qquad\qquad\qquad  \qquad\qquad \qquad\qquad
-k r^3 g_{tx_{1}}' + 2 k  \left(r^2+\omega ^2\right)g_{tx_{1}}=0\,,\nn
&2 i k \left(k^2+1\right) r^3 \omega  a_{x_{1}}+2 k^2 r^5 a_t''
+2 k^2r^4(k^2+1)  a_t'
+2 k^2 r^2 \omega ^2 a_t'
\nn
&-2 k^2 \left(k^2+1\right) r^3 a_t
-\omega ^2  \left(\left(k^2+1\right) r^2+\omega ^2\right)g_{x_{1}x_{1}}+2 i k \omega  g_{tx_{1}} \left(\left(k^2+1\right) r^2+\omega^2\right) \nn
&+2 k^2 r^5 \tau_{1} \Phi_3'   -2 r^2 \tau_{1}  \left(k^2 r^2+\omega^2\right)\Phi_3+ \left(k^4 r^4-\omega ^2 \left(r^2+\omega ^2\right)\right)\Phi_{2}+k^2 r^5 \Phi_{2}'\nn
&\qquad\qquad\qquad\qquad\qquad\qquad\qquad-r^2  \left(k^2 \left(\left(k^2+2\right) r^2+\omega ^2\right)+\omega ^2\right)\Phi_{1}=0\,,\nn
&i\omega  (g_{x_{1}x_{1}}+2 \tau_{1} \Phi_3+\Phi_{1}+\Phi_{2})+2  k \left(r^2 a_{x_{1}}'+g_{tx_{1}}\right)=0\,.
\end{align}
The first two equations comprise part of a second order system for the 
three gauge-invariant fields $\{\Phi_{1},\Phi_{2},\Phi_3\}$. We can solve the third equation for $\Phi_1'$. 
Differentiating this expression and then using the remaining equations, we find the remaining second order equation. 
Introducing the three-vector $\mathbb{V}^{T}=\left(\Phi_{1},\,\Phi_{2},\Phi_3 \right)$
we can write these in matrix form as the standard system of three mixed modes propagating on $AdS_2$:
\begin{align}
\left(\frac{\omega^{2}}{r^{2}}+r^{2}\,\partial_{r}^{2}+2r\,\partial_{r} \right)\,\mathbb{V}\,-M^{2}\,\mathbb{V}=0\,,
\end{align}
with mass matrix
\begin{align}\label{eq:mmatrix}
M^{2}=\left(\begin{array}{ccc}
2+2\tau_{1}^{2}+k^{2}   &-2k^{2} & 2\tau_{1}\left(2-k^{2}-\tau_{2}-v_{2}\right) \\
-1 &k^2  & -2\tau_{1} \\
 -\tau_{1} & 0 & k^{2}+v_{2}+\tau_{2}
\end{array}\right).
\end{align}
In addition, once we have specified these three fields the remaining four functions are determined by 
solving first order equations arising from \eqref{eq:reqns} and demanding regularity at the Poincar\'e horizon of the
$AdS_2$ space at $r=0$.

To summarise, for $D=4$ we have shown that the linearised perturbation \eqref{eq:pert} corresponds to three
propagating modes in $AdS_2$ with mass matrix \eqref{eq:mmatrix}.  We note that the mass matrix \eqref{eq:mmatrix} only depends 
on two parameters appearing in \eqref{eq:pott_exp},
$\tau_{1}$ and $\tau_{2}+v_{2}$. 
As a check we note that when $\tau_{1}=0$ the scalar field decouples from the metric and gauge-field fluctuations and we 
obtain the mass-squared eigenvalues 
\begin{align}\label{speccase}
m_{1}^{2}&=k^{2}+v_{2}+\tau_{2}\,,\nn
m^{2}_{\pm}&=1+k^{2}\pm\sqrt{1+2\,k^{2}}\,.
\end{align}
In particular, the masses $m^{2}_{\pm}$ agree with previous studies 
of longitudinal perturbations of Einstein-Maxwell theory around its $AdS_{2}\times \mathbb{R}^{2}$ solution \cite{Edalati:2010pn}. 

The $AdS_2$ BF bound is violated if any of the 
three eigenvalues $m_{i}^{2}$ of the mass matrix \eqref{eq:mmatrix} is such that $m_{i}^{2}<-1/4$.
The signal for spatially modulated phases are modes that violate the BF bound with the smallest mass-squared
occurring at $k\ne 0$. This is easily achieved by suitable choices of the parameters $\tau_1$ and $\tau_2+v_2$, 
including cases where there BF violating modes only occur for $k\ne 0$. 
Indeed, the mass squared eigenvalues when $k=0$ are simply
\begin{align}
m_{1}^{2}=0,\qquad 
m^{2}=2,\qquad m^2_{3}=2\tau_1^2+\tau_2+v_2\,,
\end{align}
and demanding that $2\tau_1^2+\tau_2+v_2\ge -1/4$ there are still broad choices for the parameters
which have modes with $k\ne 0$ that violate the BF bound. Note from \eqref{speccase} that while there can be instabilities
when $\tau_1=0$ they will not be spatially modulated.

To illustrate we can consider the specific choice of $V,\tau$ given by
\begin{align}\label{emdmodel}
V=v_0 e^{-\gamma\phi},\qquad \tau=e^{\gamma
\phi}\,,
\end{align}
where $\gamma$ is a constant, 
that were studied in \cite{Charmousis:2010zz}. The eigenvalues of the mass matrix are given by $k^2, 1+k^2\pm\sqrt{1+2(1+\gamma^2)k^2}$.
At $k=0$ eigenvalues are $0$, $0$ and $2$ and hence do not violate the BF bound. 
If $\gamma>1$ there are BF violating modes for a range of $k\ne 0$, with the minimum mass-squared occurring at
$k^2=\gamma^2(2+\gamma^2)/(2(1+\gamma^2))$.

We emphasise that the spatially modulated BF violating modes that we have identified are associated with CDWs preserving $P$ and $T$.
To see this\footnote{We will elaborate on this discussion in a simpler setting in the next section in the context of another class of models.}
we first note that the mode that is relevant for this discussion is static with $\omega=0$. From \eqref{eq:reqns} we 
deduce that these modes have $h_{tx_1}=a_{x_1}=0$ and hence the linearised perturbation in
\eqref{eq:pert} has $\delta g_{tx_1}=\delta A_{x_1}=0$. Thus, the gauge-field part of the linearised perturbation only
involves a spatially modulated $A_t$ component corresponding to a spatially modulated CDW in the dual theory. It is also
clear from the perturbation in \eqref{eq:pert} that the CDW preserves $P$ and $T$. Indeed it is worth noting  
that the mass-matrix in \eqref{eq:pert} only depends on $k^2$ in contrast to other cases that have been
analysed in the literature (e.g. equation (3.17) of \cite{Nakamura:2009tf} and equation (2.17) of \cite{Donos:2011bh}) 
where it depends linearly on $k$.

Finally, we have focussed on Einstein-Maxwell-dilaton theory in $D=4$ spacetime dimensions, but very similar comments apply to other $D$. For example,
when $D=5$, the analogue of the mass matrix given in \eqref{eq:mmatrix} is presented in appendix A.

\section{Einstein-Maxwell-dilaton-vector models}
We now consider a theory in $D$ bulk space-time dimensions which couples gravity to a massless $U(1)$ 
gauge field $A$, a scalar dilaton field $\phi$ and an additional vector field $B$ of mass $m_v$.
The Lagrangian density is given by
\begin{align}\label{eq:Lagra}
\mathcal{L}= &R -\tfrac{1}{2}(\partial \phi)^2 -V\left(\phi\right)-\tfrac{1}{4}t\left(\phi\right)
F^2-\tfrac{1}{4} v(\phi)G^2-\tfrac{1}{2}m_{v}^{2}B^2
-\tfrac{1}{2}u\left(\phi\right)FG\,,
\end{align}
where $F=dA$ and $G\equiv dB$ and $FG=F_{\mu\nu}G^{\mu\nu}$.
The corresponding equations of motion are given by
\begin{align}\label{eoms}
&R_{\mu\nu}=\tfrac{1}{D-2}V\,g_{\mu\nu}+\tfrac{1}{2}m^2_v B_{\mu}B_{\nu}+\tfrac{1}{2}\partial_{\mu}\phi\,\partial_{\nu}\phi
+\tfrac{1}{2}t \left(F_{\mu\rho}F_{\nu}{}^{\rho}-\tfrac{1}{2\left(D-2\right)}g_{\mu\nu}F_{\rho\sigma}F^{\rho\sigma} \right)\nn
&\qquad\quad+ \tfrac{1}{2}v\left(G_{\mu\rho}G_{\nu}{}^{\rho}-\tfrac{1}{2\left(D-2\right)}g_{\mu\nu}G_{\rho\sigma}G^{\rho\sigma}\right)
+u \left( G_{(\mu}{}^\rho F_{\nu)\rho}{}-\tfrac{1}{2\left(D-2\right)}g_{\mu\nu}G_{\rho\sigma}F^{\rho\sigma} \right)\, ,\nn
&\nabla_\mu\left(t\, F^{\mu\nu}+u\,G^{\mu\nu}\right)=0\, ,\nn
&\nabla_\mu\left(vG^{\mu\nu}+u\,F^{\mu\nu} \right)-\,m_{v}^{2}B^\nu=0\, ,\nn
&\nabla^2\phi-V^{\prime}-\tfrac{1}{4}t^{\prime}F^2-\tfrac{1}{4}v^{\prime}G^2-\tfrac{1}{2}u^{\prime}FG=0\, .
\end{align}

We will assume that the functions $V,t,u$ and $v$ have the following expansion
\begin{align}\label{conds}
V(\phi)&=-\frac{1}{L^2}+\tfrac{1}{2}\,m_{s}^{2}\,\phi^{2}+\cdots,\nn
t(\phi)&=1-\frac{1}{2}nL^2\,\phi^{2}+\cdots,\nn
u(\phi)&=\frac{1}{\sqrt 2}sL\,\phi+\cdots\, , \nn
v(\phi)&=1+\cdots .
\end{align}
Then $AdS_D$, with radius squared $\ell^2$, where
\begin{align}
\ell^{2}=L^2\left(D-1\right)(D-2)\,,
\end{align}
and $\phi=A=B=0$, solves the equations of motion and is dual to
a CFT in $D-1$ space-time dimensions. The gauge-field $A$ is dual to a conserved current for a
global $U(1)$ symmetry, while $\phi$ and $B$ are dual to neutral scalar and vector operators, respectively. 
In the special case that $m_v=0$ the dual CFT has a second global $U(1)$ symmetry and
$B$ is dual to the conserved current.

We are interested in the CFT held at fixed chemical potential $\mu$ with respect to the global $U(1)$ symmetry associated with
the gauge-field $A$.
The high temperature phase is described by the standard electrically charged AdS-RN black brane solution
with $\phi=B=0$.
At zero temperature, the solution interpolates between $AdS_D$ in the UV and $AdS_2\times \mathbb{R}^{D-2}$
in the IR, with the latter solution given by
\begin{align}\label{eq:ads2_bac1}
ds^{2}_{4}&=L^2\,\left(ds^{2}\left(AdS_{2}\right)+dx_{1}^{2}+\dots + dx_{D-2}^{2} \right)\,,\nn
F&=E\mathrm{vol}\left(AdS_{2}\right)\,,\nn \phi&=G=0\, .
\end{align}
where $E=\sqrt{2}L$.

In the following subsection we will first examine CDW type instabilities of this $AdS_{2}\times\mathbb{R}^{D-2}$ solution. 
Such instabilities imply that electrically charged AdS-RN will have analogous instabilities 
at finite temperature, and these will be analysed in the subsequent subsection.
We also note that the spatially modulated instabilities that we saw in the last section are not present in the models that we
consider in this section because in \eqref{conds} we have assumed $t'(0)=0$.

\subsection{Instabilities of the $AdS_{2}\times\mathbb{R}^{D-2}$ solution}
We consider the following linearised perturbation about \eqref{eq:ads2_bac1}, using
the coordinates for $AdS_2$ given in \eqref{ads2coord},
\begin{align}
\delta\phi&=e^{-i\omega t+i k x_{1}}\,\Phi(r)\,,\nn
\delta B&=e^{-i\omega t+i k x_{1}}\,\left(r^2 b_{t}(r)\,dt+i\omega\,b_{r}(r)\frac{dr}{r^{2}} \right).
\end{align}
We now substitute into the equations of motion \eqref{eoms}.
The $x_1$ component of the equation of motion for the gauge-field $B$ implies the 
constraint\footnote{Note that when $m^2_v\ne 0$ the equations of motion \eqref{eoms} imply that $\nabla_\mu \delta B^\mu=0$ and hence $\omega(b_{t}+b_{r}^{\prime})=0$. We also note that when $m^2_v=0$ we can work in a gauge with $\delta B_r=0$ if desired.}
\begin{align}
k\omega(b_{t}+b_{r}^{\prime})=0\,.
\end{align}
We also obtain a coupled system of equations which we can write in matrix form
\begin{align}
\left(\frac{\omega^{2}}{r^{2}}+r^{2}\,\partial_{r}^{2}+2r\,\partial_{r} \right)\,\mathbb{V}\,-L^{2}\,M^{2}\,\mathbb{V}=0\,,
\end{align}
where $\mathbb{V}^{T}=\left(\Phi, b_{r} \right)$ and the mass matrix is given by
\begin{align}\label{eq:mmatrix1}
M^{2}=\left(\begin{array}{cc}
m_{s}^{2}+n+s^{2}+p^{2}& -s\,(m_{v}^{2}+p^{2}) \\
-s & m_{v}^{2}+p^2 
\end{array}\right)\,
\end{align}
with $p=k/L$. 
The matrix \eqref{eq:mmatrix1} yields the $AdS_{2}$ mass spectrum
\begin{align}
m_{\pm}^{2}&=\tfrac{1}{2} (\tilde{m}_{s}^2+m_{v}^{2}+s^{2})+p^{2}\nn
   &\qquad\qquad\qquad\pm\frac{1}{2}\sqrt{\left(\tilde{m}_{s}^{2}-m_{v}^{2} \right)^{2}+2\left(\tilde{m}_{s}^{2}+m_{v}^{2}+2\,p^{2} \right)\,s^{2}+s^{4}}\,,
\end{align}
where
\begin{align}\label{emtilde}
\tilde{m}_{s}^{2}&=m_{s}^{2}+n\,.
\end{align}

It is now straightforward to choose parameters such that there are modes
violating the BF bound associated with spatially modulated phases. As a specific example,
which we will return to in the next subsection, 
if we set $D=4$, $L^2=1/24$, $m^2_s=-8$, $m^2_v=0$, 
$n=-96$, $s=16.2$ we find  $L^2 m^2_-<-1/4$ in the range 
\begin{align}\label{pval}
3.52  \lesssim |p|\lesssim 8.92\,.
\end{align}

Note that when $\omega=0$ the spatially modulated BF violating modes involves the scalar field and only the time component
of the gauge-field, $B_t$. To be more explicit, consider the static mode
 \begin{align}
\delta\phi&=e^{i k x_{1}}v_1 r^\lambda\,,\nn
\delta B&=e^{i k x_{1}}v_2 r^{\lambda+1}dt\,,
\end{align}
where $v_1,v_2$ are constants and $\lambda$ corresponds to the scaling dimension of an operator
in the one-dimensional CFT dual to the $AdS_2\times\mathbb{R}^{D-2}$ solution.
We obtain a solution to the linearised equations of motion provided that
\begin{align}
\left(
\lambda(\lambda+1)
\left(
\begin{matrix}1&0\\0&1\\
\end{matrix}
\right)\,
-L^2\left(
\begin{array}{cc}
m_{s}^{2}+n+p^{2}  & \frac{s}{L^2}(\lambda+1)\\
\lambda s  &  m_{v}^{2}+p^2
\end{array}
\right)\right)
\left(
\begin{matrix}
v_1\\v_2
\end{matrix}
\right)
=0\,.
\end{align}
This can be solved for $\lambda$ and it is straightforward to check that $\lambda(\lambda+1)=m^2_\pm$ as expected.
As usual, the BF bound is violated when $\lambda$ becomes imaginary, the onset of which happens when $\lambda=-1/2$.
These BF violating modes are associated with spatially modulated phases that preserve 
$P$ and $T$. Furthermore, in the special case that $m^2_v=0$, when there is a second global $U(1)$ symmetry in the dual CFT, the unstable modes correspond to CDWs for the second $U(1)$. We will make this more explicit in the next subsection.

\subsection{AdS-RN black hole instabilities}
In this subsection we will show, for an illustrative case, that the instabilities that we deduced for the $AdS_2\times\mathbb{R}^{D-2}$ solutions
are associated with instabilities of the finite temperature AdS-RN black brane solution. In particular, we will calculate the value of the critical temperature at which the AdS-RN black brane becomes unstable as a function of wave number, finding the usual ``bell-curve" type behaviour. 

We will work in $D=4$. The electrically charged AdS-RN black brane is given by
\begin{align}\label{adsrnsol}
ds^2&=-f dt^2+\frac{dr^2}{f}+{r^2}(dx_1^2+dx_2^2)\,,\nn
A&=(1-\frac{r_+}{r})dt\,,\nn
\phi&=B=0\,,
\end{align}
with
\begin{align}
f=\frac{r^2}{\ell^2}-\left(\frac{r_+^2}{\ell^2}+\frac{1}{4}\right)\frac{r_+}{r}+\frac{r_+^2}{4r^2}\,.
\end{align}
This describes the high temperature phase of the dual $d=3$ CFT at finite temperature $T$ and non-vanishing chemical potential $\mu$ with respect to
the global symmetry corresponding to the gauge-field $A$. Notice, for conevenience, that we have scaled to set $\mu=1$ and that 
$T=(12r_+^2-\ell^2)/(16\pi r_+\ell^2)$.
We will set $m^2_v=0$ so that the CFT has a second global
symmetry, corresponding to the gauge-field $B$. The instabilities that we discuss correspond to CDW phases associated with this second $U(1)$ symmetry.

Specifically, following from the analysis above, we consider the following linearised perturbation
\begin{align}
\delta\phi=\phi(r)\,\cos(k x_{1})\,,\nn
\delta B= b_{t}(r)\,\cos(k x_{1})dt\,.
\end{align}
For illustration we will take the scalar field to have mass given by $m_{s}^{2}=-2/\ell^{2}$. This corresponds to having an operator in the dual $d=3$ CFT with
scaling dimension $\Delta=1,2$ and we will choose the boundary conditions so that $\Delta=2$. Since we are interested in instabilities associated with
phases that spontaneously break translation invariance, we consider the following UV expansion as $r\to \infty$, in which the sources are set to zero:
\begin{align}\label{eq:UV_bc}
\phi(r)\approx \frac{\phi_{2}}{r^{2}}+\cdots,\qquad
b_{t}(r)\approx \frac{q}{r}+\cdots.
\end{align}
We will also demand regularity on the black hole horizon at $r=r_+$, by demanding that
the functions admit the analytic expansion:
\begin{align}\label{eq:nh_bc}
\phi(r)&\approx \phi_{(0)}+{\cal O}\left(r-r_{+}\right)\,,\nn
b_{t}(r)&\approx b_{(0)}\,\left(r-r_{+}\right)+{\cal O}\left(r-r_{+}\right)^{2}\,.
\end{align}

\begin{figure}[t!]
\centering
\includegraphics[width=0.6\textwidth]{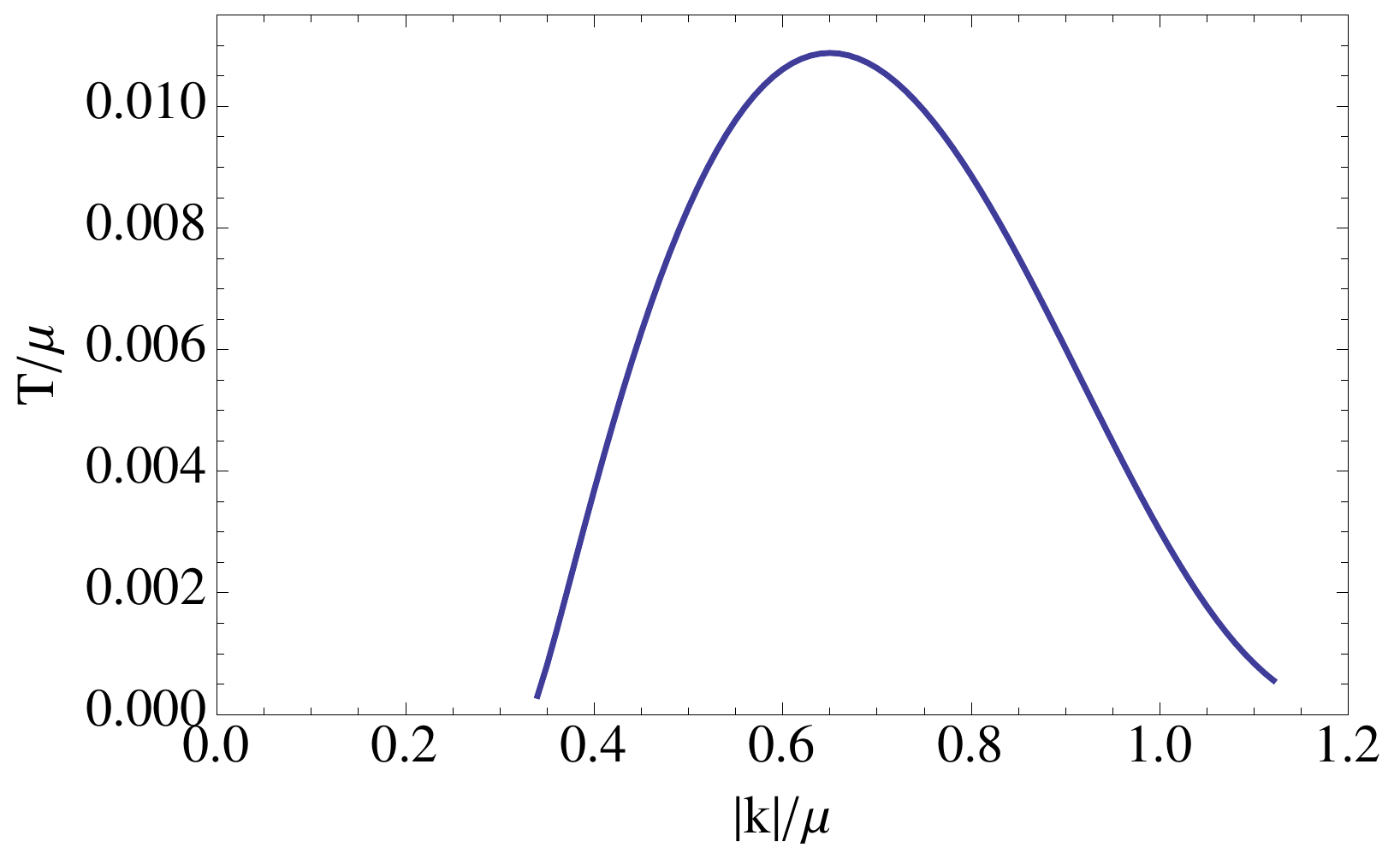}
\caption{A plot of the critical temperature versus wave number at which the electrically-charged AdS-RN black brane becomes
unstable to the formation of a CDW phase preserving $P$ and $T$. The plot is for the model \eqref{eq:Lagra}, \eqref{conds}
with $n=-96$, $s=16.2$, $m_{s}^{2}=-8$ and $m^2_v=0$. For these values the highest critical temperature occurs at 
$T_c/\mu  \approx 0.011$ with
$|k_c|/\mu \approx 0.645$.
\label{fig:Tk}}
\end{figure}

The linearised equations of motion lead to two second order differential equations for $\phi(r), b_t(r)$ and so
a solution is specified by four integration constants. For a given $k$ we have five parameters, $r_+, \phi_2,q, \phi_{(0)}, b_{(0)}$ 
entering the ODEs. Since the ODEs are linear, we can always scale one of the parameters to unity. This means that for a given $k$, we expect
solutions to exist, if at all, for specific critical values of $T$. 
These are precisely the static modes appearing at the onset of the instability.
In figure \ref{fig:Tk} we have plotted the critical temperature versus wave number for these modes
for the specific choice of parameters given by $n=-96$, $s=16.2$ and $\ell=1/2$. Recall that we used the same parameters in the last subsection and we also note that in contrast to the last subsection, the equations now depend on $m_{s}^{2}$ and $n$ individually and not
just in the combination \eqref{emtilde}. To make a comparison with the wave-numbers of the $AdS_2\times\mathbb{R}^2$ 
analysis in the last subsection given in \eqref{pval}, we should rescale the spatial coordinates by a factor of $\sqrt 2$ leading to
dividing $p$ by $4\sqrt 3$ so that \eqref{pval} corresponds to $0.508\lesssim |k|\lesssim 1.29$, in good agreement with figure \ref{fig:Tk}.

The static mode with the highest critical temperature, $T_c$,
associated with wave-number $k_c$, corresponds to the onset of CDW phases in the dual CFT. Indeed, for temperatures just below $T_c$ the charge density $j_t^{(B)}$ for the second $U(1)$, corresponding
to the gauge-field $B$, becomes spatially modulated with
\begin{align}
\langle j_t^{(B)}\rangle \sim q \cos(k_c x_1)\,.
\end{align}
At the same time the operator ${\cal O}_\phi$, dual to the scalar field $\phi$, also spontaneously acquires an expectation value given by
$\langle {\cal O}_\phi\rangle \sim \phi_2\cos(k_c x_1)$. It is clear that this new CDW phase preserves $P$ and $T$.

\section{Discussion}
We have shown that two classes of holographic models with electrically charged $AdS_2\times \mathbb{R}^{D-2}$ solutions
can have spatially modulated BF violating modes that are associated with CDWs. Specifically, these modes demonstrate the existence 
of rich classes of spatially modulated black hole solutions that are dual to the appearance of 
CDW phases, preserving both $P$ and $T$, in holographic matter at finite charge density. 
For the models discussed in section 3
with two $U(1)$ vector fields, we deduced the critical temperature at which these black hole solutions will appear.
Similar calculations are also possible for the Einstein-Maxwell-dilaton theory considered in section 2, by choosing specific
functions $V,\tau$, constructing the unbroken phase black holes and then analysing the linearised perturbations.
The CDW phases can be embedded into top down settings. For example, the model in section III is associated with Romans
theory and hence type IIB and D=11 supergravity (see the discussion in section 2 of \cite{Donos:2011qt}).

Constructing the fully back reacted black hole solutions will require solving non-linear partial
differential equations. The simplest black hole solutions will depend on two variables,
the radius $r$ and the coordinate $x_1$, being static and translationally invariant in the remaining spatial coordinates. 
However, since at the linearised level we can superimpose the static normalisable modes,
there will also be more elaborate black hole solutions that depend on all of the spatial coordinates. Only a detailed analysis
will reveal which periodic structure is thermodynamically preferred.
It would be particularly interesting to construct the black hole
solutions all the way down to zero temperature so that ground states of the system can be
identified. While it is possible that the spatial modulation disappears at zero temperature it seems
much more likely that the generic ground states will be spatially modulated CDWs. 

The results of this paper, combined with those of \cite{Donos:2012js},
 also suggest other constructions of novel spatially modulated holographic ground states.
We suppose the $AdS_2\times \mathbb{R}^{D-2}$ solution still arises as the IR limit of a domain
wall solution interpolating from some holographic behaviour in the UV. But now we assume that the 
$AdS_2\times \mathbb{R}^{D-2}$ solution is stable, without any modes violating the BF bound. 
In addition we assume\footnote{A concrete example is provide by the model \eqref{emdmodel} with $0<\gamma<1$.}
that the $AdS_2\times \mathbb{R}^{D-2}$ IR fixed point solution does not have any relevant
operators with $k=0$, but does have relevant operators with $k\ne 0$. 
If we now consider deforming the UV fixed point of the domain wall
by a spatially homogeneous and isotropic deformation, the  IR fixed point will be stable under RG flow, because
of the absence of $k=0$ relevant modes in the IR.
However, if we switch on a suitable spatially modulated deformation in the UV (such as a spatially modulated chemical potential)
the RG flow will be destabilised due to the relevant IR operators with $k\ne 0$. It seems likely that such 
constructions will also lead to rich classes of spatially modulated ground states.
As in \cite{Donos:2012js}, these constructions should include metal-insulator transitions where
the metallic phase is described by the $AdS_2\times \mathbb{R}^{D-2}$ IR  behaviour and the insulating phase
by the putative spatially modulated ground state. The spatial modulation of the charge density
suggests that such insulating phases can be interpreted as holographic Mott insulators. 

The holographic CDW phases that we have found in this paper were all in the context of models in
which the unbroken phase black holes have a zero temperature limit containing an $AdS_2$ factor in the IR.
This was purely a technical simplification and similar spatially modulated phases
should also occur in many other situations. For example, within Einstein-Maxwell-dilaton theory there can be
black holes whose zero temperature limit interpolates between some UV behaviour and a hyperscaling 
violating ground state \cite{Charmousis:2010zz,Ogawa:2011bz,Huijse:2011ef}
in the far IR and these should manifest similar phases, leading to even more spatially modulated
ground states at zero temperature. Support for the conjecture made in \cite{Donos:2011ff} 
that the generic holographic states at finite density and/or in a magnetic field are spatially modulated continues to accumulate.

\section*{Acknowledgements}
We thank Sean Hartnoll for helpful discussions.
The work is supported in part by STFC grant ST/J0003533/1.

\appendix

\section{$D=5$ mass matrix}
For Einstein-Maxwell-dilaton theory in $D$ space-time dimensions the 
conditions \eqref{eq:req}-\eqref{eq:pott_exp} are still applicable.
The relevant perturbation about the $AdS_2\times\mathbb{R}^{3}$ solution is
\begin{align}\label{eq:pertap}
\delta g_{tt}&=L^{2}r^{2}\,e^{-i\omega t} h_{tt}\left(r\right)\,\cos\left(k\,x_{1}\right)\,,\nn
\delta g_{x_{i}x_{i}}&=L^{2}\,e^{-i\omega t} h_{x_{i}x_{i}}\left(r\right)\,\cos\left(k\,x_{1}\right)\,,\nn
\delta g_{tx_{1}}&=L^{2}\,e^{-i\omega t} h_{tx_{1}}\left(r\right)\,\sin\left(k\,x_{1}\right)\,,\nn
\delta A_{t}&=-E\,e^{-i\omega t} a_{t}\left(r\right)\,\cos\left(k\,x_{1}\right)\,,\nn
\delta A_{x_{1}}&=-E\,e^{-i\omega t} a_{x_{1}}\left(r\right)\,\sin\left(k\,x_{1}\right)\,,\nn
\delta \varphi&=e^{-i\omega t} h\left(r\right)\,\cos\left(k\,x_{1}\right)\,,
\end{align}
with $h_{x_{2}x_{2}}=h_{x_{3}x_{3}}$. The analysis is very similar to that of the $D=4$ case. Using the same definition of the three scalar $AdS_{2}$ fields given in \eqref{eq:AdS2_sc_def} we obtain the mass matrix
\begin{align}\label{eq:mmatrixap}
M^{2}=\left(\begin{array}{ccc}
2+2\,\tau_{1}^{2}+k^{2}   &-4k^{2} & 2\tau_{1}\left(2-k^{2}-\tau_{2}-v_{2}\right) \\
-2/3 &k^2  & -4\tau_{1}/3 \\
 -\tau_{1} & 0 & k^{2}+v_{2}+\tau_{2}
\end{array}\right).
\end{align}


\providecommand{\href}[2]{#2}\begingroup\raggedright\endgroup

\end{document}